\def\fr#1#2{\hbox{${#1\over #2}$}}
\def\prg#1{\medskip{\bf #1}}
\def\lra{\leftrightarrow}           \def\Ra{\Rightarrow}
                  \def\pd{\partial}
\def\ort{\perp}                     
\def\mb#1{\hbox{\boldmath$#1$}}

\def\vT{{\bar T}}                   
         \def\vl{{\bar\lambda}}
                 \def\vb{{\bar b}}
\def\vg{{\bar g}}                   \def\vA{{\bar A}}
                   \def\tG{{\tilde G}}
\def\vH{\bar H}                     \def\vpi{\bar\pi}
\def\vv{\bar v}                     \def\va{\bar a}

\def\bd{{\bar\d}}

\def\dsb{\tilde b}      \def\dsl{\tilde\lambda}
\def\dsv{\tilde v}      \def\dspi{\tilde\pi}

\def\m{\mu}             \def\n{\nu}              
          \def\g{\gamma}           \def\d{\delta}
          \def\s{\sigma}           \def\t{\tau}
\def\a{\alpha}          \def\b{\beta}            \def\th{\theta}
\def\vphi{\varphi}      \def\ve{\varepsilon}     \def\p{\pi}
\def\r{\rho}            \def\D{\Delta}           \def\L{\Lambda}
\def\l{\lambda}         \def\om{\omega}          

\def\cL{{\cal L}}       \def\cH{{\cal H}}        \def\cP{{\cal P}}
\def\cM{{\cal M }}      \def\cO{{\cal O}}        \def\cE{{\cal E}}
           \def\hcO{{\hat\cO}}
\def\tP{\tilde P}       \def\tM{\tilde M}
\def\tH{\tilde H}       \def\cG{{\cal G}}

 \def\TA{\stackrel{A}{T}}
\def\tgr{\hbox{GR$_\parallel$}}       \def\grl{\hbox{GR$_\L$}}
\def\tpl{\hbox{TP$_\L$}}
\def\dfrac#1#2{\displaystyle{\frac{#1}{#2}}}
\def\nn{\nonumber}
\def\be{\begin{equation}}             \def\ee{\end{equation}}
\def\ba#1{\begin{array}{#1}}           \def\ea{\end{array}}
\def\bea{\begin{eqnarray} }           \def\eea{\end{eqnarray} }
\def\lab#1{\label{eq:#1}}             \def\eq#1{(\ref{eq:#1})}
\def\bsubeq{\begin{mathletters}}      \def\esubeq{\end{mathletters}}
\def\bitem{\begin{itemize}}           \def\eitem{\end{itemize}}

\documentstyle[aps,preprint,eqsecnum]{revtex}        

\begin{document}
\tighten

\title{Conservation laws in the teleparallel theory \\
       with a positive cosmological constant}
\author{M. Blagojevi\'c$\ ^{1}$
        and M. Vasili\'c$\ ^{2,}$\footnote{Email addresses: mb@phy.bg.ac.yu
                                        and mvasilic@phy.bg.ac.yu}}
\address{$^1$Primorska Inst. for Natural Sci. and Technology, 6000 Koper,
             P.O.Box 327, Slovenia\\
         $^2$Institute of Physics, 11001 Belgrade, P.O.Box 57, Yugoslavia}
\maketitle

\begin{abstract}
We study the conservation laws in the teleparallel theory with a
positive cosmological constant, an extension of the teleparallel theory
possessing  solutions with de Sitter asymptotics. Demanding that the
canonical generators of the asymptotic symmetry are well defined, we
obtain their improved form, which defines the conserved charges of the
theory. The physical interpretation of the results is discussed.
\end{abstract}


\section{Introduction}

The Poincar\'e gauge theory (PGT) is a gauge formulation of gravity
which represents, at least at the classical level, a viable alternative
to general relativity (GR) \cite{1,2,3}. The investigation of the
canonical structure of PGT led to a significant improvement of our
understanding of its gauge and physical properties \cite{4}. An
important conclusion following from this analysis is that a consistent
picture of the conserved charges in PGT is closely related to the
asymptotic structure of spacetime \cite{5,6,7}, in analogy to the
similar results found earlier in GR \cite{8,9}. The conclusion is based
on the study of conservation laws for the solutions which define
Minkowskian geometry in the asymptotic region. We now wish to extend
this type of the canonical analysis to a different class of solutions,
characterized by {\it de Sitter\/} (dS) asymptotic geometry. Although
some aspects of the related conservation laws have been already
discussed in the literature \cite{10}, the canonical approach will lead
us to a more clear and systematic understanding of this important
problem.

We begin our investigation of the problem in a particularly interesting
limit of PGT, defined by the {\it teleparallel\/} geometry of spacetime
\cite{11,12,13}. Basic gravitational variables in PGT are the tetrad field
$b^i{_\m}$ and the Lorentz connection $A^{ij}{_\m}$, the related field
strengths are the torsion and the curvature:
$$
T^i{}_{\m\n}=\pd_\m b^i{_\n}+ A^i{}_{m\m}b^m{_\n}-(\m\lra\n)\, ,\quad
R^{ij}{}_{\m\n}=\pd_\m A^{ij}{_\n}+A^i{}_{m\m}A^{mj}{_\n}-(\m\lra\n)\,.
$$
General geometric structure of PGT is described by Riemann-Cartan
geometry $U_4$, characterized by a metric compatible connection. In the
PGT context, the teleparallel (Weitzenb\"ock) geometry $T_4$ is defined by
the requirement of vanishing curvature:
\be
R^{ij}{}_{\m\n}(A)=0\, .                                  \lab{1.1}
\ee
The observational relevance of the teleparallel theory (TP) is based on
the existence of the one-parameter family of Lagrangians, which is
empirically indistinguishable from GR. For a specific value of the
parameter, the Lagrangian of the theory coincides, modulo a divergence,
with the Einstein-Hilbert Lagrangian, and defines \tgr, the
teleparallel form of GR \cite{11,12,13}.

Many important features of TP have been investigated in the literature,
including its gauge structure, the canonical formulation, the initial
value problem, and the form of various differential identities
\cite{14,15,16,17,18}. Moreover, the teleparallel approach was
successfully applied in studying the tensorial proof of the positivity
of energy in GR, the transparent treatment of Ashtekar's complex
variables, and the construction of the teleparallel Kaluza-Klein theory
\cite{19,20,21}. In the present paper we introduce the teleparallel
theory with cosmological constant (\tpl), an extension of TP  which
posses solutions with the dS asymptotics. The conservation laws in
\tpl\ are then investigated by generalizing our earlier results
\cite{22}, related to the class of solutions in TP with Minkowskian
asymptotics.

The value of the cosmological constant $\L$ compatible with the
observational data is extremely small, perhaps even zero, and leads to
the fine tuning problem in the standard model of fundamental
interactions \cite{23}. The investigation of the consistency of
theories with $\L\ne 0$ may help us to better understand this unusual
phenomenon. Thus, for instance, in GR with cosmological constant
(\grl), the perturbative stability of the ground state has been  studied
using a suitably defined concept of the Killing energy \cite{24}. The
present paper lays the grounds for such consistency tests in \tpl\
\cite{25}.

The layout of the paper is as follows. In Sec. II we introduce the
Lagrangian for the teleparallel theory with cosmological constant,
and analyze some particular solution of the field equations. We use
these results in Sec. III to choose the vacuum solution of \tpl,
analyze its symmetries and construct the related global symmetry
generator.  The vacuum solution is characterized by the tetrad
field which defines the dS metric, and the non-vanishing vector
component of the torsion. Its symmetry is found to be a subgroup of
$SO(1,4)$ --- the 3-dimensional Weyl group $W(3)$. We choose the
asymptotic behaviour of all phase-space variables in Sec. IV, and
use the Regge-Teitelboim method \cite{9} to define the improved form
of the global generators in Sec. V. The Hamiltonian conservation
laws are presented in Sec. VI, and finally, both physical and
geometric meaning of the results is discussed in Sec. VII. Some
technical details are given in the Appendix.

Our conventions are the same as in Refs. \cite{16,22}.

\section{The teleparallel theory with a cosmological constant}

\subsection{General structure}

In the framework of PGT, the general teleparallel theory with a
cosmological constant is described by the Lagrangian density
\bea
&&\cL = b\bigl(\cL_T -2a\L\bigr)
  +\l_{ij}{}^{\m\n}R^{ij}{}_{\m\n}+b\cL_M\, ,\nn\\
&&\cL_T=a\left(AT_{ijk}T^{ijk}+BT_{ijk}T^{jik}+CT_k T^k\right)
       \equiv \b_{ijk}(T)T^{ijk}\, .                      \lab{2.1}
\eea
Here, $\l_{ij}{}^{\m\n}$ are Lagrange multipliers  introduced to
ensure the teleparallelism condition \eq{1.1}, $\L$ is the
cosmological constant, $a=1/16\p G$ ($G$ is Newton's gravitational
constant), $A,B$ and $C$ are free parameters, $T_k=T^m{}_{mk}$, and
$\cL_M$ is the matter field Lagrangian.

By varying the Lagrangian \eq{2.1} with respect to $b^i{_\m}$,
$A^{ij}{_\m}$ and $\l_{ij}{}^{\m\n}$, we obtain the gravitational
field equations:
\bsubeq\lab{2.2}
\bea
&&4\nabla_\r\bigl(b\b_i{^{\m\r}}\bigr)
  -4b\b^{nm\m}T_{nmi}+h_i{^\m}b(\cL_T-2a\L) =\t^\m{_i}\, ,\lab{2.2a}\\
&&4\nabla_\r\l_{ij}{}^{\m\r}-8b\b_{[ij]}{}^\m=\s^\m{}_{ij}\, ,
                                                          \lab{2.2b}\\
&& R^{ij}{}_{\m\n}=0\, ,                                  \lab{2.2c}
\eea
\esubeq
where $\t^\m{_i}$ and $\s^\m{}_{ij}$ are the energy-momentum and spin
currents of matter fields, respectively.

The physical relevance of the teleparallel theory with $\L=0$ is based
on the fact that there is a one-parameter subclass of this theory,
defined by
$i)$ $2A+B+C=0$, $C=-1$,
which is empirically indistinguishable from GR \cite{12,13}. In
particular, for the parameter value
$ii)$ $B=1/2\,$ (i.e. $A=1/4, B=1/2, C=-1$),
the one-parameter theory goes over into the so-called teleparallel
formulation of GR \cite{15}, the formulation in which the
gravitational field equations coincide with Einstein's equations (at
least for scalar matter). After having clarified the gauge structure
and conservation laws in the teleparallel theory with $\L=0$ \cite{22},
we now turn our attention to the one-parameter teleparallel theory with
$\L\ne 0$.

\subsection{The one-parameter \tpl}

We begin our study of the one-parameter \tpl\ by analyzing some
interesting properties of its field equations, which will be useful in
defining the vacuum solution of the theory. The expressions for
$\b_{ijk}$ and $\cL_T$ in the one-parameter theory read:
\bea
&& \b_{ijk}=\b_{ijk}^\parallel-\frac{a}{4}(2B-1)\TA_{ijk}\, ,\nn\\
&& \cL_T=\cL_T^{\parallel}-\frac{a}{12}(2B-1)\TA_{ijk}\TA{}^{ijk}\,,\nn
\eea
where $\b_{ijk}^\parallel$ and $\cL_T^\parallel$ are the \tgr\
expressions, and $\TA_{ijk}=T_{ijk}+T_{kij}+T_{jki}$. Using the
identity \cite{13}
$$
abR(A)=abR(\D)+b\cL_T^\parallel -\pd_\r(2abT^\r)\, ,
$$
where $R(\D)$ is the Riemannian scalar curvature, the Lagrangian of
the one-parameter \tpl\ can be written in the form
\be
\cL = -ab\bigl[R(\D)+2\L\bigr]
   -\frac{a}{12}(2B-1)b\TA_{ijk}\TA{}^{ijk}
   +\dsl_{ij}{}^{\m\n}R^{ij}{}_{\m\n}+b\cL_M +{\rm div}\, ,\lab{2.3}
\ee
with
$\dsl_{ij}{}^{\m\n}\equiv\l_{ij}{}^{\m\n}+aH_{ij}^{\m\n}/2$, and
$H_{ij}^{\m\n}=b(h_i{^\m}h_j{^\n}-h_j{^\m}h_i{^\n})$.
The first two field equations take the form:
\bsubeq\lab{2.4}
\bea
&&R_{ik}(\D)-\fr{1}{2}\eta_{ik}\bigl[R(\D)+2\L\bigr]
          +\cO(\TA)=\t_{ki}/2ab\, ,                  \lab{2.4a}\\
&&\nabla_\r\left( \l_{ij}{}^{\m\r}+\frac{a}{2}H_{ij}^{\m\r} \right)
          +\cO(\TA)=\s^\m{}_{ij}/4\, ,               \lab{2.4b}
\eea
\esubeq
where $\cO(\TA)$ denotes terms proportional to $\TA_{ijk}$.
It is now simple to conclude that
for the field configurations satisfying $\TA_{ijk}=0$,
{\it the first field equation has the same form as in \grl\/}.
The consistency of this equation requires
$\t_{ki}$ to be symmetric. Taking into account that the second
field equation serves only to determine the Lagrange multipliers
$\l_{ij}{}^{\m\n}$, we can use this conclusion to generate some
solutions of \tpl, starting from certain solutions of \grl.
Consider, for instance, a metric which has diagonal form in some
coordinate system:
\bsubeq
\be
ds^2=A(dx^0)^2-B_1(dx^1)^2-B_2(dx^2)^2-B_3(dx^3)^2\, .     \lab{2.5a}
\ee
Let us choose the tetrad components to be diagonal,
\be
b^0{}_0=\sqrt{A}\, ,\qquad b^a{_\a}=\d^a_\a\sqrt{B_\a}\, , \lab{2.5b}
\ee
\esubeq
and fix the gauge $A^{ij}{}_\m=0$. Then, one easily proves that
$\TA_{ijk}=0$, and derives an important consequence:
\bitem
\item[] If the diagonal metric \eq{2.5a} is a solution
of \grl, the related tetrad \eq{2.5b} is a solution of the
one-parameter \tpl, in the gauge $A^{ij}{_\m}=0$ and with the same
$\t_{ik}$.
\eitem
An important class of solutions of this type is the class of
spherically symmetric solutions \cite{12,26}.

\subsection{Schwarzschild-de Sitter solution of \tpl}

The spherically symmetric solution of \grl\ in the region without
matter ($\t_{ik}=0$) is known as the Schwarzschild-de Sitter (SdS)
solution. In the isotropic coordinates $(t,x^1,x^2,x^3)$, it is
defined by the diagonal metric
\be
A=\left(1-\psi\over 1+\psi\right)^2\, ,
  \qquad B_1=B_2=B_3=f^2\left(1+\psi\right)^4\, ,          \lab{2.6}
\ee
where
$$
\psi=mG/2\r f\, ,\quad \r^2=\mb{x}^2\, ,\quad
f=\exp(t/\ell)\, ,\quad \ell^2=3/\L>0\, .
$$
The gravitational field is determined by the mass parameter $m$.
For $m=0$, the metric \eq{2.6} reduces to the dS form (Appendix A).

Let us now consider the one-parameter \tpl\ in the region where
matter is absent ($\t_{ik},\s^\m{}_{ik}=0$). If we fix the gauge
$A^{ij}{_\m}=0$, the preceding discussion implies that the diagonal
tetrad field
\bsubeq\lab{2.7}
\be
\tilde b^i{}_\m\equiv
\left( \ba{cc}
       \dfrac{1-\psi}{1+\psi}  & 0  \\
             0 &   f(1+\psi)^2\d^a_\a
       \ea  \right) \, ,                                 \lab{2.7a}
\ee
is a solution of the first field equation \eq{2.4a}. In addition
to that, we display here the related solution for $\l_{ij}{}^{\m\n}$:
\be
\dsl_{ij}{}^{\m\n}=-\frac{a}{2}\tilde H_{ij}^{\m\n}\, .   \lab{2.7b}
\ee
\esubeq
It is easily derived as a particular solution of the second field
equation \eq{2.4b}; the general solution differs from \eq{2.7b} by
an arbitrary lambda gauge transformation.

The SdS solution \eq{2.7} plays an important role in our study of
the vacuum structure of \tpl.

\section{The vacuum structure of \tpl}

We begin our discussion of the conservation laws in the one-parameter
\tpl\ by clarifying the vacuum structure of the phase space. An
overview of the canonical structure of \tpl, which is similar but not
the same as the one of TP, is given in Appendix B.

\subsection{The dS vacuum}

For $m=0$, the SdS solution of \grl\ reduces to the dS solution \eq{A.2},
\be
ds^2=dt^2-f^2\bigl[(dx^1)^2 + (dx^1)^2 + (dx^1)^2 \bigr]
    =\bar g_{\m\n}dx^\m dx^\n\, .                          \lab{3.1}
\ee
which defines, in a manner discussed above, a solution of \tpl.
{\it We choose this particular solution to be the vacuum of \tpl\/}.
Thus, the vacuum values of $b^i{_\m}$ and $A^{ij}{}_\m$ have the form
\bsubeq\lab{3.2}
\be
\vb^i{}_\m=\left( \ba{cc}
               1 &      0 \\
               0 &      f\d^a_\a \\
                  \ea  \right) \, ,
\qquad \vA^{ij}{}_\m = 0\, ,                               \lab{3.2a}
\ee
while the vacuum value of $\l_{ij}{}^{\m\n}$ is given by
\be
\vl_{ij}{}^{\m\n}=-\frac{a}{2}\vH_{ij}^{\m\n} \, ,         \lab{3.2b}
\ee
in accordance with \eq{2.7}.

The vacuum values of the Lagrangian variables, in conjunction with
the relations defining the canonical momenta, lead to:
\bea
&&\vpi_i{^0},\vpi_{ij}{^0},\vpi^{ij}{}_{\m\n}=0\, ,\nn\\
&&\vpi_0{^\g}=0\, ,\qquad
  \vpi_a{^\g}=-\frac{4a}{\ell}f^2\d_a^\g \, ,\nn\\
&&\vpi_{a0}{}^\a=2af^2\d_a^\a\, ,\qquad
  \vpi_{ab}{}^\a=0\, .                                    \lab{3.2c}
\eea
\esubeq

Similarly, using the Hamiltonian field equations, we can determine
the vacuum values of the Hamiltonian multipliers.

It is important to observe that our vacuum solution \eq{3.2}  is {\it time
dependent\/}. This unusual feature will have significant influence on the
canonical structure of the theory.

We now wish to clarify certain geometric aspects of the dS vacuum in the
one-parameter \tpl. It is clear that the relation
$\vb^i{_\m}\vb_{i\n}=\vg_{\m\n}$ does not fix the vacuum value of
$b^i{_\m}$ uniquely, but only up to an arbitrary local ``Lorentz"
transformation $L(b)$ that acts {\it exclusively\/} on the tetrads (this
transformation should be clearly distinguished from the true Lorentz
transformation, which acts on all the fields in the theory). Thus, to the
dS vacuum solution $\vg_{\m\n}$ of \grl, there corresponds a whole class of
solutions $\vb^i{_\m}$ of \tpl, the elements of which are related to each
other by $L(b)$. Since $L(b)$ is not a gauge symmetry of the theory, all
these solutions are {\it gauge inequivalent}. In certain sense, our choice
\eq{3.2} is the simplest one. To characterize this simplicity more
precisely, we note that the only nonvanishing vacuum components of the
torsion are
$$
\vT^a{}_{0\g}=\frac{1}{\ell}\d^a_\g f\, .
$$
If we define the irreducible torsion components as
\bea
&&v_i=T^s{}_{si}\, ,\qquad a_i=\frac{1}{3}\ve_{ijkl}T^{jkl}\, ,\nn\\
&&t_{ijk}=T_{(ij)k}
  +\frac{1}{3}\eta_{k(i}v_{j)}-\frac{1}{3}\eta_{ij}v_k\, ,\nn
\eea
we easily verify that their vacuum values are
$$
\bar v_i=-\frac{3}{\ell}\d^0_i\, ,\qquad\bar a_i= \bar t_{ijk}=0\, .
$$
Thus, our vacuum solution is characterized by the following two
conditions:
\bea
&& a)\quad \vb^i{_\m}\vb_{i\n}=\bar  g_{\m\n}\, , \nn\\
&& b)\quad \vv_i=-\frac{3}{\ell}\d^0_i\, ,
     \quad \va_i=\bar t_{ijk}=0\, .                       \lab{3.3}
\eea
The simplicity of our choice \eq{3.2} is expressed by the geometric
property that only one component of the torsion, $\bar v_i$, is
different from zero.

The existence of the privileged direction $\bar v_i$ means that the
vacuum \eq{3.2} of \tpl\ (the teleparallel geometry) is not isotropic,
in spite of the fact that the dS vacuum \eq{3.1} of \grl\ (Riemannian
geometry) is maximally symmetric (isotropic and homogeneous). More
details about the vacuum symmetries will be given in the next
subsection.

In the vacuum \eq{3.2}, we have not only $\bar a_i=0$ but also $\bar
t_{ijk}=0$. Hence, we can generalize the one-parameter Lagrangian
\eq{2.3} by adding the $t_{ijk}$ term, and still have the same vacuum
solution \eq{3.2}. The resulting two-parameter theory is defined by
only one condition, $2A+B+3C=0$, which ensures the absence of the $v_i$
term in the action. Nevertheless, our further discussion will be
restricted to the one-parameter \tpl.

\subsection{Symmetries of the dS vacuum}

The action \eq{2.1} of the general teleparallel theory is invariant
under the Poincar\'e  gauge transformations \cite{16}. The dS vacuum
\eq{3.2} is invariant only under a subgroup of these transformations
--- the subgroup that leaves the vacuum configuration invariant. It
is defined by demanding the following conditions on the gauge
parameters:
\bea
&&\om^i{_s}\vb^s{_\m}
   -\xi^\r{}_{,\m}\,\vb^i{_\r}-\xi^\r\vb^i{}_{\m,\r}=0\, ,\nn\\
&&\pd_\m\om^{ij}=0\, ,         \nn\\
&&\om_i{^s}\vl_{sj}{}^{\m\n} +\om_j{^s}\vl_{is}{}^{\m\n}
  +\xi^\m{}_{,\r}\vl_{ij}{}^{\r\n} +\xi^\n{}_{,\r}\vl_{ij}{}^{\m\r}
  -\pd_\r(\xi^\r\vl_{ij}{}^{\m\n}) =0\, .                 \lab{3.4}
\eea
The solutions of the first two equations for $\xi^\m$ and $\om^{ij}$
are of the form
\bea
&&\xi^0=\ve^0\, ,\qquad
    \xi^\a=\ve^\a+\ve^\a{}_\b x^\b -\ve^0\frac{1}{\ell}x^\a\, ,\nn\\
&&\om^0{}_b=0\, ,\qquad \om^a{}_b=\ve^a{}_b\, ,           \lab{3.5}
\eea
where $\ve^\m$ and $\ve^a{_b}=-\ve_b{^a}$ are constants, and
$\ve^\a{_\b}\equiv \d_a^\a\d_\b^b\ve^a{_b}$
( $\ve^{\a\b}=\eta^{\b\g}\ve^\a{_\g}$ ).
With these values for
$\xi$ and $\om$, the last equation in \eq{3.4} is trivially satisfied.
The global symmetry transformations of dynamical variables are
obtained from their gauge transformations, derived in Ref. \cite{16},
by replacing the gauge parameters with the vacuum values \eq{3.5}. In
this way we obtain
\bea
\bar\d b^a{_0}&=&\ve^a{_c}b^c{_0}-\ve^0 T_0 b^a{_0}
               -(\ve^\g+\ve^\g{_\b}x^\b)\pd_\g b^a{_0}\, ,\nn\\
\bar\d b^a{_\a}&=&\ve^a{_c}b^c{_\a}+\ve_\a{^\b}b^a{_\b}
                +\ve^0 \left(\frac{1}{\ell}-T_0\right)b^a{_\a}
                -(\ve^\g+\ve^\g{_\b}x^\b)\pd_\g b^a{_\a}\, ,\nn\\
\bar\d\p_0{^\a}&=&\ve^\a{_\b}\pi_0{^\b}
   +\ve^0 \left(\frac{2}{\ell}-T_0\right)\pi_0{^\a}
   -(\ve^\g+\ve^\g{_\b}x^\b)\pd_\g\pi_0{^\a} \, ,         \nn\\
\bar\d\p_a{^\a}&=&\ve_a{^c}\pi_c{^\a}+\ve^\a{_\b}\pi_a{^\b}
   +\ve^0 \left(\frac{2}{\ell}-T_0\right)\pi_a{^\a}
   -(\ve^\g+\ve^\g{_\b}x^\b)\pd_\g\pi_a{^\a} \, ,         \nn\\
\bar\d\pi_{ij}{^\a}&=&\ve_i{^c}\pi_{cj}{^\a}+\ve_j{^c}\pi_{ic}{^\a}
  +\ve^\a{_\g}\pi_{ij}{^\g}+\ve^0\left(\frac{2}{\ell}-T_0\right)
  -(\ve^\g+\ve^\g{_\b}x^\b)\pd_\g\pi_{ij}{^\a}\, ,       \lab{3.6}
\eea
where $T_0\equiv \pd_0-(x^\g/\ell)\pd_\g$, and similarly for the
other variables.

The vacuum symmetry is specified by the seven parameters:
$\ve^0,\ve^\a$, and $\ve^\a{}_\b$. The Killing vector field
$\xi=\xi^\m\pd_\m$ has the form
\bea
\xi\,&&=\ve^0\left(\pd_0-\frac{1}{\ell}x^\b\pd_\b\right)
       +(\ve^\a+\ve^\a{}_\b x^\b)\pd_\a   \nn\\
  &&\equiv\ve^0 T_0+\ve^\a T_\a-\fr{1}{2}\ve^{\a\b}L_{\a\b}\,,\nn
\eea
where $x_\a=\eta_{\a\b}x^\b$. The Lie algebra of the global
generators reads:
\bea
&&[T_0,T_\a]=\frac{1}{\ell}T_\a\, ,\qquad [T_\a,T_\b]=0\, ,\nn\\
&&[L_{\a\b},T_0]=0\, ,\qquad
  [L_{\a\b},T_\g]=\eta_{\b\g}T_\a-\eta_{\a\g}T_\b\, ,\nn\\
&&[L_{\a\b},L_{\g\d}]=\eta_{\b\g}L_{\a\d}
                      -\eta_{\a\g}L_{\b\d}-(\g\lra\d)\,.  \lab{3.7}
\eea
Thus, the isometry group of the dS vacuum is not $SO(1,4)$, as one might
naively expect, but rather the {\it subgroup\/} that leaves the vacuum
value of $v_i$ unchanged. According to the Lie algebra \eq{3.7}, this
subgroup coincides with the three-dimensional Weyl group $W(3)$.

Since the vacuum solution depends on time, the simple time translation,
generated by the vector $\ve^0\pd_0$, is {\it not\/} the vacuum symmetry.

\subsection{The global symmetry generator}

The canonical generator of the local Poincar\'e transformations in
\tpl\ is found in Appendix B.  The global generator $G$,  corresponding
to the vacuum symmetry of the theory, is obtained by substituting
the vacuum values \eq{3.5} for $\xi$ and $\om$  into the expression
for the local Poincar\'e generator. Using the result \eq{B.2}, we find
\bsubeq\lab{3.8}
\be
G=-\ve^\m P_\m+\fr{1}{2}\ve^{\a\b}M_{\a\b}\, ,             \lab{3.8a}
\ee
where
\bea
&&P_0=\int d^3x \left( \cP_0-\frac{1}{\ell}x^\a\cP_\a\right)\, ,
  \qquad P_\a=\int d^3x \cP_\a\, ,   \nn\\
&&M_{\a\b}=\int d^3x\cM_{\a\b}\,,\qquad
  \cM_{\a\b}=x_\a\cP_\b- x_\b\cP_\a-S_{\a\b}\, ,\         \lab{3.8b}
\eea
\esubeq
The quantities $\cP_\m$ and $S_{\a\b}=S_{ab}\d^a_\a\d^b_\b$ are
defined in Appendix B. The action of the generator $G$ on a phase
space functional $X$ is given by $\bar\d X=\{X,G\}$.

\section{Asymptotic structure of spacetime}

In this section we discuss the behaviour of the dynamical variables at
large distances, where they approach their vacuum values \eq{3.2}. We
begin these considerations by identifying a set of basic (minimal)
asymptotic conditions, which will be necessary for a consistent
treatment of the asymptotic dynamics. Then, as a complement to these,
we derive the asymptotic form of the Hamiltonian constraints.
These two sets of asymptotic requirements will prove to be sufficient
to determine the form of the conservation laws.

\subsection{Asymptotic conditions}

The dS space, treated as a Riemannian space, is the maximally symmetric
spacetime with positive cosmological constant. Its spatial sections,
determined by $t=$ const, are 3-spheres which have no
boundaries, hence there is no notion of spatial infinity. For this
reason, it is not simple to define an asymptotically dS spacetime. In
our approach, we rely on the analogy with the SdS form of the metric
and its behaviour for $\psi\to 0$. We say that a given metric $g$ is
asymptotically dS if
\bitem
\item[] $g_{\m\n}=\vg_{\m\n}+\cO_1\,$ at large $\r\, .$
\eitem
This property  defines {\it finite\/} gravitational sources,
characterized by the absence of matter at large $\r$, where spacetime
becomes nearly dS.

Thus, we are going to identify the region of large $\r$ as the asymptotic
region \cite{27}. This is an acceptable approach if we treat our theory as
an {\it ordinary field theory\/}, the dynamical variables of which are well
defined for all values of $\r$. It will enable us to formally apply the
complete canonical apparatus to the field theory \eq{2.3}, and derive the
related conservation laws. However, the {\it geometric interpretation\/} of
this procedure will have serious limitations related to the existence of
the cosmological horizon in the (vacuum) dS geometry, as discussed in Sect.
VII.

Continuing with our field-theoretic approach, we now wish to determine
the behaviour of the tetrad fields for large $\r$. One could start with
the general assumption $b^i{_\m}=\bar b^i{_\m}+\cO_1$. However, for
spherically symmetric solutions corresponding to finite sources, the
first order corrections to the vacuum solution are of the SdS type.
Moreover, studying the effects of the spinning matter Hayashi and
Shirafuji \cite{12} showed that the related corrections are of the
second order. Thus, even in this case, the first order terms are of the
SdS type. In GR, this property is always fulfilled, provided we chose
convenient gauge conditions. The results of the post-Newtonian analysis
of the one-parameter TP theory indicate that this might be true also
for \tpl\ \cite{12,13}, but an independent analysis is necessary to
prove this.

Motivated by these considerations, we adopt the following asymptotic
conditions:
\bsubeq\lab{4.1}
\be
b^i{_\m}=\dsb^i{_\m}+\cO_2\, ,\qquad A^{ij}{_\m}=\hcO\, ,\lab{4.1a}
\ee
where $\dsb^i{_\m}$ is the SdS tetrad field \eq{2.7a}, the second relation
is compatible with the gauge character of $A^{ij}{_\m}$, and $\hcO$ denotes
a term with arbitrarily fast asymptotic decrease. The asymptotic behaviour
of the $\l$ variable is assumed to have the form
\be
\l_{ij}{}^{\m\n}=\vl_{ij}^{\m\n} +\cO_1\, ,             \lab{4.1b}
\ee
in accordance with the field equation \eq{2.2b}. In addition to this, the
matter fields, being highly localized (finite gravitational sources), are
chosen to have arbitrarily fast decrease.

There is a general principle which we use to define the asymptotic
behaviour of dynamical variables: the {\it expressions which vanish on
shell should have an arbitrarily fast asymptotic decrease\/}, as no
solution of the field equations is thereby lost.
In particular, the constraints of the theory are assumed to decrease
arbitrarily fast, and consequently, all volume integrals defining the
global generators \eq{3.7} are convergent. In this way, the
relations that define the canonical momenta lead to the following
asymptotics:
\bea
&& \pi_i{^0},\pi_{ij}{^0},\pi^{ij}{}_{\m\n}=\hcO\, ,\nn\\
&&\pi_0{^\g}=\cO_2\, ,\qquad
  \pi_a{^\g}=\dspi\d_a^\g +\cO_2\, ,\nn\\
&&\pi_{a0}{}^\a=2af^2\d_a^\a+\cO_1\, ,\qquad
  \pi_{ab}{}^\a=\cO_1\, .                               \lab{4.1c}
\eea
\esubeq
Here, and in what follows, we use the notation
\bea
&&\dsb^a{_\a}=\dsv\d^a_\a\, ,\qquad \dspi_a{^\a}=\dspi\d_a^\a\,,\nn\\
&&\dsv=f(1+\psi)^2\, ,\qquad \dspi=-\frac{4a}{\ell}\dsv^2\, .\nn
\eea

In the present analysis, we shall extensively use the property of spatial
derivatives to enhance asymptotic decrease: $\pd_\a\cO_n=\cO_{n+1}$. This
property follows from the assumption that {\it there are no gravitational
waves in the asymptotic region\/}, i.e. we consider {\it isolated\/}
gravitational systems for a limited period of time.

\subsection{Asymptotic form of the constraints}

Starting with the minimal asymptotic requirements \eq{4.1}, we now turn
our attention to the form of the sure constraints in the asymptotic
region. The asymptotic form of $\cH_\a$, $\cH_{ij}$ and $\cH_\ort$ is
given by the following relations (Appendix C):
\bsubeq\lab{4.2}
\bea
\cH_\a&&=-\pd_\g\cP_\a{^\g}+\cO_5\, ,                  \lab{4.2a}\\
\cH_{\a\b}&&=\cP_{\a\b}-\cP_{\b\a}
             +\pd_\g\pi_{\a\b}{^\g}+\cO_4 \, ,         \lab{4.2b}\\
\cH_{0\b}&&=f\left( \pi_{0\b}+\frac{4a}{\ell}fb_{0\b}\right)
            +\pd_\g\pi_{0\b}{^\g} +\cO_3\, ,           \lab{4.2c}\\
\cH_\ort&&=\frac{1}{\ell}\cP_\a{^\a}-\pd_\b\pi_0{^\b}+\cO_4\, ,
                                                       \lab{4.2d}
\eea
\esubeq
where
\be
\cP_\a{^\g} \equiv b^a{_\a}\pi_a{^\g}+\frac{4a}{\ell}J\d_\a^\g\,.
                                                       \lab{4.3}
\ee
In these relations, $\cH_{\m\n}\equiv\d^i_\m\d^j_\n\cH_{ij}$,
$\pi_{\m\n}{^\r}\equiv \d^i_\m\d^j_\n\pi_{ij}{^\r}$,
$\cP_{\a\b}\equiv\eta_{\b\g}\cP_\a{^\g}$ and
$\pi_{i\b}\equiv \eta_{\b\g}\pi_i{^\g}$.
The asymptotics of $\cH^{ij}{}_{\a\b}$ does not yield any new
information, while $\phi_{ij}{^\a}=\pi_{ij}{^\a}-4\l_{ij}{}^{0\a}$
defines the asymptotic behaviour of $\pi_{ij}{^\a}$.

\section{Improving the global generators}

In the Hamiltonian theory, the generators of symmetry
transformations act on dynamical variables via the Poisson bracket
operation, which is defined in terms of functional derivatives. A
functional
$F[\vphi,\p]=\int d^3x f(\vphi,\pd_\m\vphi,\p,\pd_\n\p)$
has well defined functional derivatives if its variation can be
written in the form
\be
\d F=\int d^3x\bigl[ A(x)\d\vphi(x)+B(x)\d\p(x)\bigr]\,,  \lab{5.1}
\ee
where terms $\d\vphi_{,\m}$ and $\d\p_{,\n}$ are absent. In
addition, the well defined phase-space functionals have to be finite
on their whole domain.

The global symmetry generators \eq{3.8} do not satisfy these requirements,
although, when acting on local expressions, they produce the correct
transformation laws. Our idea is to improve their form by adding
appropriate surface terms, so that they can also act on global phase-space
functionals. This would be a very useful achievement, since it would allow
us to calculate their Poisson brackets, laying thereby the ground for our
discussion of the conservation laws in the next section.

\prg{Linear momentum.} We begin our discussion with the spatial translation
generator $P_\a=\int d^3x\cP_\a$. Its variation is given by \cite{22}
\bsubeq\lab{5.2}
\bea
\d\cP_\a&&=-\d (b^i{_\a}\p_i{^\g})_{,\g}
           +(\p_i{^\g}\d b^i{_\g})_{,\a}+R+\pd\hcO\nn\\
        &&=-\d (b^a{_\a}\p_a{^\g})_{,\g}
           +(\p_a{^\g}\d b^a{_\g})_{,\a}+R+\pd\cO_4\, ,        \lab{5.2a}
\eea
where $R$ denotes terms of the regular form \eq{5.1}.
The second term in $\d\cP_\a$ can be transformed using the idea of
decomposing any dynamical variable into its SdS part and
the rest. This is possible owing to the adopted asymptotics
\eq{4.1}. Introducing the notation
$$
\D b^a{_\g}\equiv b^a{_\g}-\dsb^a{_\g}\, ,
$$
and similarly for the other variables, one can show that
\bea
\p_a{^\g}\d b^a{_\g}&&=\dspi_a{^\g}\d\dsb^a{_\g}
  +\dspi_a{^\g}\d\D b^a{_\g}+(\D\pi_a{^\g})\d\dsb^a{_\g}+\cO_4\nn\\
&&=-\frac{4a}{\ell}\d(\dsv^3+\dsv^2\D b^a{_a})
   +\left(\D\pi_a{^a}+\frac{8a}{\ell}f\D b^a{_a}\right)\d\dsv
   +\cO_4\nn\\
&&=-\frac{4a}{\ell}\d J+\cO_4\, , \nn
\eea
where the constraint $\cH_\ort\approx 0$ is used in the form
\eq{C.6}. Consequently,
\be
\d\cP_\a=-\d\cP_\a{^\g}{}_{\!,\g}+R+\pd\cO_4\, ,          \lab{5.2b}
\ee
\esubeq
and the improved linear momentum takes the form
\bea
&&\tP_\a=P_\a+E_\a\, ,\nn\\
&&E_\a=\oint dS_\g\cP_\a{^\g} \, .                         \lab{5.3}
\eea
The constraint $\cH_\a\approx 0$ ensures that the surface integral
$E_\a$ is finite.

\prg{Angular momentum.} The variation of $M_{\a\b}$ has
the form
\bea
&&\d M_{\a\b}=\int d^3 x\d\cM_{\a\b}\, ,\nn\\
&&\d\cM_{\a\b}=x_\a\d\cP_\b -x_\b\d\cP_\a+\d\cH_{\a\b}+R+\hcO\, .\nn
\eea
Using the expression \eq{5.2b} for $\d\cP_\a$, and the relation
$\d\cH_{\a\b}=\d\pi_{\a\b}{^\g}{}_{,\g}+R$, we obtain
\bea
&& \tM_{\a\b}=M_{\a\b}+E_{\a\b}\, ,\nn\\
&&E_{\a\b}=\oint dS_\g\left(x_\a\cP_\b{^\g}-x_\b\cP_\a{^\g}
                    -\pi_{\a\b}{^\g}\right) \, .         \lab{5.4}
\eea
The constraints $\cH_\a\approx 0$ and $\cH_{\a\b}\approx 0$ imply
that $E_{\a\b}$ is finite:
\bea
&&\pd_\g\left(x_\a\cP_\b{^\g}-x_\b\cP_\a{^\g}
                             -\pi_{\a\b}{^\g}\right)\nn\\
&&=-\left(\cP_{\a\b}-\cP_{\b\a}+\pd_\g\pi_{\a\b}{}^\g\right)
   +\cO_4=\cO_4\, .\nn
\eea

\prg{Killing energy.} To find the improved form of $P_0$, the global
symmetry generator which produces translations along the Killing time
vector $T_0^\m=(1, -x^\a/\ell)$, we begin with
$$
\d P_0=\int d^3x\left(\d\cP_0-\frac{x^\a}{\ell}\d\cP_\a\right)\,.
$$
The expression for $\d\cP_0$ can be found from the relations \cite{22}
\bsubeq\lab{5.5}
\bea
&&\d\cP_0=-\d(b^k{_0}\p_k{^\g})_{,\g}+\d\cH_T +\pd\hcO\, , \nn\\
&&\d\cH_T\approx -\pd_\a\left( {\pd\cL\over \pd b^k{}_{\m,\a}}
                 \d b^k{_\m}\right)+\pd\hcO+R\, ,        \lab{5.5a}
\eea
The first term in $\d\cH_T$ is a 3-divergence of
$L^\a\equiv-4b\b_{ijk}h^{j\a}h^{k\m}\d b^i{_\m}$.
Taking into account the adopted asymptotics, we find
$$
L^\a= \pi_a{^\a}\d b^a{_0}+\cO_3
    = \d\bigl( \pi_a{^\a} b^a{_0} \bigr)+\cO_3 \, ,
$$
where the last equality follows from $b^a{_0}\d\pi_a{^\a}=\cO_3$.
Thus,
\bea
\d\cP_0&&=-\d(b^k{_0}\p_k{^\g})_{,\g}+\d(b^a{_0}\p_a{^\g})_{,\g}
   +R+\pd\cO_3 \nn\\
       &&=-\d(b^0{_0}\p_0{^\g})_{,\g}+R+\pd\cO_3\, .     \lab{5.5b}
\eea
\esubeq
This result, together with the relation
$$
x^\a\d\cP_\a=-\d\left(x^\a\cP_\a{^\g}\right)_{,\g}+R+\pd\cO_3\, ,
$$
leads to the improved form of $P_0$:
\bea
&&\tP_0=P_0+E_0\, ,\nn\\
&&E_0=\oint dS_\g\left(\pi_0{^\g}
                 -\frac{x^\a}{\ell}\cP_\a{^\g}\right)\,.  \lab{5.6}
\eea
The finiteness follows from the constraints $\cH_\a\approx 0$ and
$\cH_\ort\approx 0$:
$$
\pd_\g \left(\pi_0{^\g}-\frac{x^\a}{\ell}\cP_\a{^\g}\right)=
\left(\pd_\g\pi_0{^\g}-\frac{1}{\ell}\cP_\a{^\a}\right)+\cO_4=\cO_4\, .
$$

\prg{SdS charges.} As a preliminary test that our expressions for the
gravitational charges $E_\a, E_{\a\b}$ and $E_0$ are meaningful, we
evaluate their values for the SdS solution. Using the constraint
$\cH_{0\b}\approx 0$ and the known form of the SdS solution, one finds
$$
\tilde\pi_0{^\a}=\frac{m}{4\pi}\frac{n^\a}{\r^2}+\cO_3\, .
$$
This result, together with $\tilde\cP_\a{^\g}=0$, leads to
$\tilde E_0=m$. We also find that the SdS values of the linear and
angular momentum vanish: $\tilde E_\a=0$, $\tilde E_{\a\b}=0$.

\section{Conservation laws}

Once we have the improved symmetry generators, it is not difficult to show
their on-shell conservation, provided the time evolution is governed by the
well defined Hamiltonian. The necessary and sufficient conditions for a
well defined phase-space functional $\tG[\vphi,\p,t]$ to be a generator of
global symmetries take the form \cite{22}
\be
\frac{\pd\tilde G}{\pd t}+ \left\{\tG\,,\tH\right\}=C_{PFC}\ , \qquad
\left\{\tilde G\,,\phi_m\right\}\approx 0\, ,                  \lab{6.1}
\ee
where $C_{PFC}$ is a primary first class constraint, and $\phi_m$ are all
the constraints in the theory. The strong equality in the above relations
is equality up to trivial generators; these can be, for instance, squares
of constraints or pure constants (central terms). The structure of these
relations has to be in agreement with the general result of Ref. \cite{28},
stating that the Poisson bracket of two well defined generators is
necessarily a well defined generator.
Comparing  the time evolution equation for $\tG$ with the first condition
in \eq{6.1}, we see that our symmetry generator is conserved only when the
central terms are absent.

In standard situations with {\it static vacuums\/}, the canonical
temporal evolution leaves the vacuum invariant, and consequently,
the Hamiltonian generates the global time translation symmetry of
the vacuum. However, our present choice of the canonical time is
such that we have a {\it time dependent vacuum\/}, which is not
invariant under the action of $\tH$. As a consequence, the domain of
the finite and differentiable functional $\tH$, as defined by our
asymptotic conditions \eq{4.1}, is not preserved during the time
evolution of the system, and therefore, strictly speaking, $\tH$ is
{\it not\/} a well defined evolution generator. This unsatisfactory
situation can be handled in either of two ways: a) we can extend our
phase space by allowing a suitable class of vacuum configurations to
replace our single vacuum state, or b) we could try to obtain the
conservation laws working with well defined symmetry generators
only. In this case, we find it much simpler to follow the second
approach, which is equivalent to \eq{6.1}, but technically
different. It is based on the fact that the Hamiltonian is
essentially a part of the well defined global generator $\tP_0$, so
that all the conservation laws are contained in the Poisson bracket
algebra between $\tP_0$ and the global generators
$\tG=(\tP_\m,\tM_{\a\b})$.

To define this approach, we observe that our improved generators
have the form of integrals of some local densities: $\tP_0=\int d^3x
(\cG_0+\cE_0)$, $\tG=\int d^3x(\cG+\cE)$, where $\cG$'s are
constraints. The Poisson bracket $\{\tP_0,\tG\}$ can be calculated
in two equivalent, but technically different ways:

i) by acting with $\tG$ on the integrand of $\tP_0$:
\bsubeq
\be
\{\tP_0,\tG\}=
  \int d^3x\{\cG_0+\cE_0,\tG\}\approx\int d^3x\{\cE_0,\tG\}
             \propto\int d^3x\bd_G\cE_0 \, ;             \lab{6.2a}
\ee

ii) by acting with $\tP_0$ on the integrand of $\tG$:
\be
\{\tP_0,\tG\}=\int d^3x\{\tP_0,\cG+\cE\} \approx
    \int d^3x\{\tP_0,\cE\}\propto \int d^3x\bd_1\cE\, .   \lab{6.2b}
\ee
\esubeq
Here, the weak equality appears after observing that $\cG_0$ and
$\cG$, being the constraints, ``commute" with all symmetry
generators. In the first procedure, $\bd_G\cE_0$ is the $\tG$
transformation of $\cE_0$ which, according to the known global
transformation rules \eq{3.6}, {\it does not\/} contain the time
derivative of $\cE_0$. On the other hand, the expression $\bd_1\cE$
in the second procedure is the $\tP_0$ transformation of $\cE$ which
{\it does\/} involve the time derivative of $\cE$. The equality of
the two results gives an information on the time derivative of
$\tG\approx\int d^3x\cE$, which will help us to find whether
the generator $\tG$ is conserved or not.

To realize the two procedures just described, we need the transformation
laws \eq{3.6} and the relation
$$
\bar\d\cP_\a{^\g}=\ve_\a{^\b}\cP_\b{^\g}+\ve^\g{_\b}\cP_\a{^\b}
      +\ve^0\left(\frac{3}{\ell}-T_0\right)
      -(\ve^\d+\ve^\d{_\b}x^\b)\pd_\d\cP_\a{^\g}\, ,
$$
which is easily derived from \eq{3.6}. We also introduce the simplifying
notation
$$
\bd_1=\bd(\ve^0)\, ,\qquad \bd_2=\bd(\ve^\a)\, ,
                    \qquad \bd_3=\bd(\ve^{\a\b})\, ,
$$
for the global transformations with parameters $\ve^0$, $\ve^\a$ and
$\ve^{\a\b}$, respectively.

\subsection{Poisson bracket algebra}

The first step in our procedure is to
calculate $\{\tP_0,\tG\}$ using equation \eq{6.2a}. We expect to obtain
relations representing the canonical realization of the part of the Weyl
algebra \eq{3.7} involving $\tP_0$.
The first result is rather trivial:
\bsubeq\lab{6.3}
\be
\{\tP_0,\tP_0\}=0\, .                                         \lab{6.3a}
\ee

Continuing with $\tG=\ve^\a\tP_\a$, we find
$$
\{\tP_0,\ve^\a\tP_\a\}\approx -\oint dS_\g
         \bd_2\left(\pi_0{^\g}-\frac{x^\a}{\ell}\cP_\a{^\g}\right)\, .
$$
When restricted to the $\ve^\a$ parameter, the needed transformation
laws are given by
$$
\bar\d_2\pi_0{^\g}= -\ve^\b\pd_\b\pi_0{^\g}\, ,\qquad
\bar\d_2\cP_\a{^\g}=-\ve^\b\pd_\b\cP_\a{^\g}\, .
$$
Using the relations $\pd_\b\pi_0{^\g}=\cO_3$ and
\bea
-\pd_\g\left(x^\a\pd_\b\cP_\a{^\g}\right)&=&
  -\pd_\b\left(\cP_\g{^\g}+x^\a\pd_\g\cP_\a{^\g}\right)
  +\pd_\g\cP_\b{^\g} \nn\\
  &=& \pd_\g\cP_\b{^\g}+\pd\cO_3\, ,\nn
\eea
we obtain
\be
\{\tP_0,\tP_\a\}\approx \frac{1}{\ell}E_\a
                \approx \frac{1}{\ell}\tP_\a\, .           \lab{6.3b}
\ee

Now, we turn our attention to $\tG=\ve^{\a\b}\tM_{\a\b}/2$:
$$
\left\{\tP_0,\frac{1}{2}\ve^{\a\b}\tM_{\a\b}\right\}\approx
\oint dS_\g\bd_3\left(\pi_0{^\g}-\frac{x^\a}{\ell}\cP_\a{^\g}\right)\,,
$$
where
\bea
&&\bd_3\pi_0{^\g}=\ve^\g{_\b}\pi_0{^\b}
                   -\ve^\d{_\b}x^\b\pd_\d\pi_0{^\g}\, ,\nn\\
&&\bd_3\cP_\a{^\g}=\ve_\a{^\b}\cP_\b{^\g}+\ve^\g{_\b}\cP_\a{^\b}
                   -\ve^\d{_\b}x^\b\pd_\d\cP_\a{^\g}\, .\nn
\eea
Using the identities
\bea
&&\pd_\g\bd_3\pi_0{^\g}=
    -\pd_\b\left[\ve^\b{_\d}x^\d(\pd_\g\pi_0{^\g})\right]\,, \nn\\
&&\pd_\g\bd_3(x^\a\cP_\a{^\g})=
      -\pd_\b\left[\ve^\b{_\d}x^\d(\cP_\g{^\g}
      +x^\a\pd_\g\cP_\a{^\g})\right]\,,         \nn
\eea
and the constraints \eq{4.2a} and \eq{4.2b}, we find that
$\pd_\g\bd_3\left(\pi_0{^\g}-x^\a\cP_\a{^\g}/\ell\right)=\pd\cO_3$,
and consequently, the surface integral vanishes. Therefore,
\be
\{ \tM_{\a\b},\tP_0\}\approx 0\, .                        \lab{6.3c}
\ee
\esubeq

The results \eq{6.3} represent the part of the Poisson bracket algebra of
the global generators that involves $\tP_0$, and is in complete agreement
with \eq{3.7}. The same kind of calculations would give us the rest of the
algebra, but we don't need it here.

\subsection{Conserved charges}

We now perform similar calculations based on the second procedure defined
in \eq{6.2b}, with the conclusion that our gravitational charges are
conserved.

\prg{Killing energy.} For $\tG=\tP_0$, we have
$$
\{\ve^0\tP_0,\tP_0\}\approx -\oint dS_\g
   \bd_1\left(\pi_0{^\g}-\frac{x^\a}{\ell}\cP_\a{^\g}\right)\, .
$$
The needed transformation rules are
\bea
&&-\bd_1\pi_0{^\g}=\ve^0\pd_0\pi_0{^\g}
    +\frac{1}{\ell}\ve^0\left[
     \pi_0{^\g} -\pd_\b(x^\b\pi_0{^\g})\right]\, , \nn\\
&&\bd_1\left(x^\a\cP_\a{^\g}\right)=
    -\ve^0{x^\a}\pd_0\cP_\a{^\g}
    +\frac{1}{\ell}\ve^0\left[
     {x^\a}\pd_\b\left({x^\b}\cP_\a{^\g}\right)\right]\, . \nn
\eea
Then, after combining the identities
\bea
&&\pd_\g\left[ \pi_0{^\g}-\pd_\b(x^\b\pi_0{^\g})\right]=
  -\pd_\b(x^\b\pd_\g\pi_0{^\g})\, ,  \nn\\
&&\pd_\g \left[x^\a\pd_\b(x^\b\cP_\a{^\g})\right]=
  \pd_\b(x^\b\cP_\g{^\g})+\pd_\b(x^\a x^\b\pd_\g\cP_\a{^\g})  \nn
\eea
with the constraints \eq{4.2a} and \eq{4.2d}, we obtain
$$
-\oint dS_\g\bd_1\left(\pi_0{^\g}-\frac{x^\a}{\ell}\cP_\a{^\g}\right)=
 \ve^0\pd_0\oint dS_\g\left(\pi_0{^\g}
                       -\frac{x^\a}{\ell}\cP_\a{^\g}\right)\, .
$$
Therefore,
\bsubeq
\be
\{\tP_0,\tP_0\}\approx\dot E_0 \, ,                        \lab{6.4a}
\ee
and the relation \eq{6.3a} implies that the Killing energy is conserved:
\be
\dot E_0\approx 0\, .                                      \lab{6.4b}
\ee
\esubeq

\prg{Linear momentum.} For $\tG=\tP_\a$ in \eq{6.2b}, we have
\bea
&&\{ \ve^0\tP_0,\tP_\a\}\approx \oint dS_\g \bd_1\cP_\a{^\g}\, ,\nn\\
&&\bd_1\cP_\a{^\g}=-\ve^0\pd_0\cP_\a{^\g}
                  +\frac{1}{\ell}\ve^0\pd_\b(x^\b\cP_\a{^\g})\,.\nn
\eea
Now, the identity
$$
\pd_\g\left[\pd_\b(x^\b\cP_\a{^\g})\right]=\pd_\g\cP_\a{^\g}+
   \pd_\b(x^\b\pd_\g\cP_\a{^\g})=\pd_\g\cP_\a{^\g}+\pd\cO_4
$$
leads to
\bsubeq
\be
\{\tP_0,\tP_\a\}\approx -\dot E_\a+\frac{1}{\ell}E_\a\, . \lab{6.5a}
\ee
This relation together with \eq{6.3b} implies the conservation of
the linear momentum:
\be
\dot E_\a\approx 0\, .                                    \lab{6.5b}
\ee
\esubeq

\prg{Angular momentum.} To verify the conservation of $\tM_{\a\b}$,
we start with
$$
\{\ve^0\tP_0,\tM_{\a\b}\}\approx \oint dS_\g
 \bd_1\left(x_\a\cP_\b{^\g}-x_\b\cP_\a{^\g}-\pi_{\a\b}{}^\g\right)\, ,
$$
where
\bea
&&\bd_1(x_\a\cP_\b{^\g})-(\a\lra\b)=
   \ve^0 x_\a\left[-\pd_0\cP_\b{^\g}
  +\frac{1}{\ell}\pd_\d(x^\d\cP_\b{^\g})\right]-(\a\lra\b)\, ,\nn\\
&&\bd_1\pi_{\a\b}{}^\g=\ve^0\left[-\pd_0\pi_{\a\b}{}^\g
  +\frac{1}{\ell}(2+ x^\d\pd_\d)\pi_{\a\b}{}^\g\right] \, .\nn
\eea
We transform these expressions using the identities
\bea
&&\pd_\g\left[ x_\a\pd_\d(x^\d\cP_\b{^\g})\right]-(\a\lra\b)=
   -\pd_\d\left[ x^\d(\cP_{\a\b}-\cP_{\b\a})\right]
   +\pd\cO_3\, ,                                             \nn\\
&&-\pd_\g\left[(2+ x^\d\pd_\d)\pi_{\a\b}{}^\g \right]=
  -\pd_\d(x^\d\pd_\g\pi_{\a\b}{}^\g)\, ,\nn
\eea
where the first equality follows with the help of \eq{4.2a}. As a
consequence of the constraint \eq{4.2b}, the sum of these two terms takes
the trivial value $\pd\cO_3$, so that
\bsubeq
\be
\{\tP_0,\tM_{\a\b}\}\approx -\dot E_{\a\b}\, .                \lab{6.6a}
\ee
Then, the angular momentum conservation easily follows from
\eq{6.3c}:
\be
\dot E_{\a\b}\approx 0\, .                                    \lab{6.6b}
\ee
\esubeq

\subsection{Lagrangian formalism}

After having discussed the canonical conservation laws, we now derive the
Lagrangian form of the gravitational charges.
A straightforward but lengthy calculation (Appendix D) leads to the
following relations:
\bsubeq
\bea
&&\cP_\a{^\g}=\,\hat h_\a{}^{0\g} + \cO_3\, ,    \nn\\
&&\p_0{^\g}-\frac{x^\a}{\ell}\cP_\a{^\g}
   =\,\hat h^{00\g}-\frac{x^\a}{\ell}\,\hat h_\a{}^{0\g}
    +\,\cO^{[\a\g]}{}_{,\,\a} + \cO_3\, ,                   \lab{6.7a}
\eea
where $\cO^{[\a\g]}$ stands for an antisymmetric quantity, whose
divergence gives no contribution to the surface integral. The
quantity $\hat h^{\m 0\l}$ is defined by
\bea
&&\hat h^{\m 0\l}\equiv \frac{1}{\sqrt{-g}}\,h^{\m 0\l}
 +\frac{4a}{\ell}\,J\left(g^{\m\l}-\frac{1}{2}\,g^{0\m}g^{0\l}\right)
 -a(2B-1)J\TA{^{\m 0\l}} \, ,                \nn\\
&& h^{\m\n\l}\equiv \pd_\r\psi^{\m\n\l\r}\,,\qquad
\psi^{\m\n\l\r}\equiv a(-g)
\left(g^{\m\n}g^{\l\r}-g^{\m\l}g^{\n\r}\right)\,,           \lab{6.7b}
\eea
\esubeq
and its indices are lowered by the true metric $g_{\m\n}$. Thus, all
terms except $\TA{^{\m 0\l}}$ are expressed through the metric
($J^2=-gg^{00}+\cO_4$). These formulas enable us to express our
canonical gravitational charges in the Lagrangian form:
\bea
&&E_\a = \oint dS_\g\,\hat h_\a{}^{0\g}\, , \nn\\
&&E_0 = \oint dS_\g\left(\hat h^{00\g}
      -\frac{x^\a}{\ell}\,\hat h_\a{}^{0\g}\right)\, ,\nn\\
&&E_{\a\b}=\oint dS_\g\left[ x_\a\hat h_\b{}^{0\g}
      -x_\b\hat h_\a{}^{0\g}+\frac{f}{(-g)}\psi^\g{}_{\a\b 0}
      -4\left(\l_{\a\b}{}^{0\g}+\frac{a}{2}H^{0\g}_{\a\b}\right)
       \right]\, .                                        \lab{6.8}
\eea

In order to compare these results with the related calculations in the
teleparallel theory with $\L=0$ \cite{22}, one should consider the limit
$\ell\to\infty$. Note that the asymptotic conditions are also changed in
this limit. Effectively,  the required transition in \eq{6.8} is achieved
by making the replacements
$$
\frac{1}{\ell}\to 0\, ,\quad \frac{f}{(-g)}\to 1\, ,\quad
\hat h^{\m 0\l}\to h^{\m 0\l} -a(2B-1)J\TA{^{\m 0\l}} \, ,
$$
whereupon we find a complete agreement with Ref. \cite{22}.

An alternative form of the Lagrangian formulas \eq{6.8} can be found by
introducing a different notation: $g_{\a\b}=f^2(\eta_{\a\b}+G_{\a\b})$,
$g_{00}=1+G_{00}$. Thus, for instance, the linear momentum in the limit
$2B-1=0$ (\tgr) has the form
\bea
E_\a=af^3\oint dS_\g\left[ \frac{2}{\ell}
      \left( G^0{_0}\d_\a^\g-G_\a{^\g}\right)
    +f^{-2}\pd_0(f^2G_\a{^\g}) -\d_\a^\g \pd_0G^\b{_\b}
       \right]   \, ,        \nn
\eea
in agreement with the related implicitly given result in Ref. \cite{24}.
Similar formulas can be derived also for $E_0$ and $E_{\a\b}$.

\section{Concluding remarks}

Here, we wish to comment on some geometric aspects of our treatment
of the conservation laws in the teleparallel theory with the dS
ground state.

\prg{1.} It is an attractive property of the Killing time vector that it
defines the symmetry of the vacuum. However, $T_0$ is not timelike
everywhere, but only in the region where
\be
T_0^2=1-f^2\r^2/\ell^2<0\, ,\quad
            \hbox{i.e.}\quad  f^2\r^2<\ell^2\, ,            \lab{7.1}
\ee
which is the interior of the horizon of the dS space (Appendix A).
Therefore, the Killing time can be interpreted as the physical time only
{\it within the horizon\/}. As a consequence, $\tP_0$ can be interpreted as
the physical energy of the system only within the horizon. The same is true
for all the other global symmetry generators.

\prg{2.} When the dS radius $\ell$ is large, the value of the Killing
energy is very close to its GR value. In spite of that, the geometric
restriction to the inner part of the horizon has serious consequences.
In our field-theoretic canonical approach developed so far, we worked in
the region of space defined by $0\le\r<\infty$. In particular, the
Hamiltonian and all the symmetry generators are defined in this region.
However, the true geometric distance is measured by $r=f\r$, so that, if we
restrict our discussion to the region $r\le L$ contained in the interior of
the horizon, $\r$ must be restricted to
\be
\r\le\frac{L}{f} \quad (L<\ell) \, .                        \lab{7.2}
\ee
This condition has an essential influence on the conservation laws. Namely,
our conserved charges are given as integrals over the whole region
$0\le\r<\infty$, and consequently, the restriction \eq{7.2} spoils their
conservation. The bigger the dS radius $\ell$, the smaller is the error we
make. Thus, the restriction \eq{7.2} implies that the very concept of
measurable energy, as well as its numerical value and the conservation law,
are defined only {\it approximately\/}. This is true for all symmetry
generators, and represents a distinctive feature of the dS space
\cite{24}.

\prg{3.} For finite $\ell$, our expressions for the gravitational charges
are exactly conserved, but they do not have a definite physical
interpretation in the whole spacetime.  On the other hand, after we impose
the restriction \eq{7.2} which ensures a clear physical interpretation, the
charges are conserved only approximately. In the limit $\ell\to\infty$, the
gravitational charges reduce to the form found earlier in the teleparallel
theory with an $M_4$ asymptotics. In this limit, they are both exactly
conserved and have a clear physical interpretation.

\prg{4.} In conclusion, the results obtained in the present paper lead to a
clear understanding of the conditions under which the gravitational charges
in the one-parameter teleparallel theory with the dS vacuum, are conserved.

\acknowledgements

This work was supported by the Serbian Science foundation, Yugoslavia, and
by the Slovenian Science foundation, Slovenia. One of us (M.B.) would like
to thank Dj. \v Sija\v cki for a discussion concerning the meaning of the
Lie algebra \eq{3.7}.

\newpage
\appendix
\section{On de Sitter geometry}

In Riemannian geometry, de Sitter (dS) space is maximally symmetric
spacetime with negative curvature (positive cosmological constant)
\cite{29}. It can be simply realized as a 4-dimensional hypersphere
$$
H_4:\quad (y^0)^2-(y^1)^2-(y^2)^2-(y^3)^2-z^2=-\ell^2
$$
embedded in a 5-dimensional Minkowski space $M_5$ with metric
$\eta=(+,-,-,-;-)$. It has the topology $R^1\times S^3$,
the signature $(+,-,-,-)$, and negative curvature, $R=-12/\ell^2$.
By construction, dS space has the isometry group $SO(1,4)$, and the
Killing vectors in coordinates $y^M=(y^\m,z)$ are
$K_{MN}=y_M\pd_N-y_N\pd_M$ (pseudo rotations in $M_5$).

In dS space, one can not introduce the standard concept of energy,
since there is no Killing vector field which is timelike everywhere.
While the rotations in $M_5$ are spacelike,  boosts are timelike only
in some regions of the dS space. Thus, for instance, the boost
$K_{05}=(-z,0,0,0,y^0)$ is timelike only for
$\mid z\mid >\mid y^0\mid$. This is an essential property of the dS
space, which is related to the existence of the (cosmological) horizon,
associated to a given observer. The best we can do is to introduce the
concept of energy within the horizon.

{\bf a)} In the pseudo-spherical coordinates
$$
\begin{array}{l}
  y^0=\ell\sinh t'\, ,\\
  z=\ell\cosh t'\cos\chi\, , \\
  \phantom{x}
\end{array} \qquad
        \begin{array}{l}
          y^1=\ell\cosh t' \sin\chi \sin\th \cos\vphi\, ,\\
          y^2=\ell\cosh t'\sin\chi\sin\th\sin\vphi\nn\, ,\\
          y^3=\ell\cosh t'\sin\chi\cos\th\, ,
        \end{array}
$$
the metric on $H_4$ becomes
\be
ds^2=\ell^2d t'^2-\ell^2\cosh^2 t'\bigl[ d\chi^2+\sin^2\chi
                      (d\th^2+\sin^2\th d\vphi^2)\bigr]\, .   \lab{A.1}
\ee
Except for the trivial singularities at $\chi=0,\pi$ and $\th=0,\pi$,
these coordinates cover the whole dS space. The space can be visualized
as a hyperboloid (take $\chi=\th=\pi/2$, for simplicity). The smallest
$S^3$ is at $t'=0$; in the future or past, 3-spheres expand
exponentially.

{\bf b)} On $H_4$ one can also introduce the coordinates
$$
\t=\ell\ln{y^0+z\over\ell}\, ,\quad x^\a={\ell y^\a\over y^0+z}\, ,
$$
in which the metric takes the form
\be
ds^2=d\t^2 -f^2\bigl[ (dx^1)^2 +(dx^1)^2 +(dx^1)^2 \bigr]\, ,
  \qquad f\equiv e^{\t/\ell}\, .                             \lab{A.2}
\ee
These coordinates cover half of the hypersphere $H_4$, since
$\t$ is defined only for $y^0+z>0$.

{\bf c)}  Let us now introduce new coordinates,
$$
\begin{array}{l}
y^0=\ell\cos u\sinh t\, ,\\
z=\ell\cos u\cosh t\, ,\\
\phantom{x}
\end{array} \qquad
       \begin{array}{l}
        y^1=\ell\sin u \sin\th \cos\vphi\, ,\\
        y^2=\ell\sin u\sin\th\sin\vphi\nn\, ,\\
        y^3=\ell\sin u\cos\th\, ,
        \end{array}
$$
which lead to
$$
ds^2=\ell^2\cos^2u dt^2-\ell^2du^2
                       -\ell^2\sin^2u(d\th^2+\sin^2\th d\vphi^2)\, .
$$
By the coordinate transformation $r=\ell\sin u$, we obtain the line
element in the static (Schwarzschild) coordinates:
\be
ds^2=\ell^2\left( 1-r^2/\ell^2\right)dt^2
-{dr^2\over{1-r^2/\ell^2}}-r^2\bigl(d\th^2+\sin^2\th d\vphi^2\bigr)\, .
                                                             \lab{A.3}
\ee
In these coordinates, the singularity of the metric components at
$r=\ell$ is clearly visible; it can be interpreted as a cosmological
singularity surrounding an observer at $r=0$.

Solving the Killing equations for the dS metric \eq{A.2} one finds the
set of ten Killing vectors. In particular, the Killing vector
$T_0=\pd_0-(1/\ell)x^\a\pd_\a$ describes the translation of the
time coordinate $\t$ followed by a dilatation of the spatial
coordinates $x^\a$. This vector is timelike within the horizon:
\be
(T_0)^2=1-{f^2\r^2\over\ell^2}>0\quad\Ra\quad f^2\r^2<\ell^2\,.\lab{A.4}
\ee
Note that the metric components \eq{A.2} are not singular at the
horizon, while those in \eq{A.3} are singular at the horizon $r=\ell$,
where $T_0$ becomes light-like.

\section{Hamiltonian and gauge generators}

The Hamiltonian structure of the general TP theory with $\L=0$ is
described in Appendix A of Ref. \cite{22}. Here, we discuss
modifications necessary for the transition TP $\to$ \tpl.

The only change in the structure of the \tpl\ Hamiltonian comes from
the presence of the cosmological constant term $-2ab\L$ in the
Lagrangian \eq{2.1}. Hence, it is simple to conclude the following:
\bitem
\item[i)] the primary constraints in \tpl\ are the same as in TP;
\item[ii)] the new canonical Hamiltonian is obtained by the replacement
          $\cH_c\to\cH_c+2ab\L$;
\item[iii)] all the sure (i.e. always present) secondary constraints,
with the exception of  $\cH_\ort$, remain of the same form;
\item[iv)] there are no sure tertiary constraints (since the Poincar\'e
and $\l$ gauge generators in \tpl\ do not posses parameters with second
time derivatives, as will be clear soon);
\item[v)] all the sure determined multipliers are unchanged (the
related second class constraints
$\phi_{ij}{}^\a=\pi_{ij}{}^\a-4\l_{ij}{}^{0\a}$ and $\pi^{ij}{}_{0\b}$
``commute" with the correction term $2ab\L$ in $\cH_c$).
\eitem

The only change in the structure of the Hamiltonian is given by
the change of the dynamical component of $\cH_c$:
\be
\cH_\ort\to\cH_\ort+2aJ\L\, .                              \lab{B.1}
\ee
Consequently, the form of the Poincar\'e gauge generator in \tpl\
is obtained from the corresponding generator in TP by the replacement
\eq{B.1}. Thus, we have:
\bsubeq\lab{B.2}
\be
G=G(\om)+G(\xi)\, ,                                       \lab{B.2a}
\ee
where
\bea
G(\om)=&&-\fr{1}{2}\dot\om^{ij}\p_{ij}{^0}
             -\fr{1}{2}\om^{ij}S_{ij}\, ,\nn\\
G(\xi)=&&\,-\dot\xi^0\bigl(
   b^k{_0}\p_k{^0}+\fr{1}{2}A^{ij}{_0}\p_{ij}{^0}
  +\fr{1}{4}\l_{ij}{}^{\a\b}\p^{ij}{}_{\a\b} \bigr)-\xi^0\cP_0 \nn\\
&&-\dot\xi^\a\bigl(
   b^k{_\a}\p_k{^0}+\fr{1}{2}A^{ij}{_\a}\p_{ij}{^0}
  -\fr{1}{2}\l_{ij}{}^{0\b}\p^{ij}{}_{\a\b}\bigr)-\xi^\a \cP_\a\, .
                                                          \lab{B.2b}
\eea
In the above expressions, the omitted integration over a
3-dimensional spatial hypersurface is always understood, and we use
the following notation:
\bea
S_{ij}=&&-\cH_{ij}+ 2b_{[i0}\p_{j]}{^0}
 +2A^s{}_{[i0}\p_{sj]}{^0}+2\l_{s[i}{}^{\a\b}\p^s{}_{j]\a\b}\, ,\nn\\
\cP_0\equiv &&\hat\cH_T=\cH_T-\pd_\a D^\a\, ,\nn\\
\cP_\a= &&\cH_\a-\fr{1}{2}A^{ij}{_\a}\cH_{ij}
 +2\l_{ij}{}^{0\b}\cH^{ij}{}_{\a\b} +\p_k{^0}\pd_\a b^k{_0}
 +\fr{1}{2}\p_{ij}{^0}\pd_\a A^{ij}{_0}\nn\\
        &&-\fr{1}{4}\l_{ij}{}^{\b\g}\pd_\a\p^{ij}{}_{\b\g}
 -\fr{1}{2}\pd_\g\bigl(\l_{ij}{}^{\b\g}\p^{ij}{}_{\a\b}\bigr)\, .
                                                         \lab{B.2c}
\eea
with
\bea
&&\cH_\a=\pi_i{^\b}T^i{_{\a\b}}-b^k{_\a}\nabla_\b\pi_k{^\b}
     +\fr{1}{2}\pi^{ij}{}_{0\a}\nabla_\b\l_{ij}{}^{0\b}\, ,\nn\\
&&\cH_{ij}=2\pi_{[i}{^\a}b_{j]\a}+\nabla_\a\pi_{ij}{^\a}
     +2\pi^s{}_{[i0\a}\l_{sj]}{}^{0\a}\, ,\nn\\
&&\cH^{ij}{}_{\a\b}=R^{ij}{}_{\a\b}
     -\fr{1}{2}\nabla_{[\a}\pi^{ij}{}_{0\b]}\, ,  \nn\\
&&D^\a=b^k{_0}\pi_k{^\a}+\fr{1}{2}A^{ij}{_0}\pi_{ij}{^\a}
     -\fr{1}{2}\l_{ij}{}^{\a\b}\pi^{ij}{}_{\a\b} \, .    \lab{B.2d}
\eea
\esubeq

The standard Poincar\'e gauge transformations of fields, given in
Eq. (2.3) of Ref. \cite{16}, are the symmetry transformations  of \tpl.
Following the same procedure as described in Ref. [CQG], one can verify
that the generator \eq{B.2a} produces the correct transformation rules
of all the fields and momenta. [The derivation is based on the following
two observations:
a) $\cH_T$ does not depend on the derivatives of momenta, and
b) $\dot Q=\{Q,H_T\}$ for any dynamical variable $Q$.]
Hence, the above expression for $G$ is the correct generator of
Poincar\'e gauge transformations in the general \tpl\ theory.

\section{Asymptotic form of the constraints}

In this Appendix we calculate the asymptotic form of the constraints
with an accuracy which is sufficient for studying the structure of the
surface terms.

Using $N=b^0{_0}+\cO_4$ and the orthogonality relations for the tetrads,
one can derive the following useful formulae:
\bea
&&b=b^0{_0}J+\cO_4\, ,\qquad \D J=\dsv^2\D b^a{_a}+\cO_4\, ,\nn\\
&&\D b^0{_0}+(\tilde b^0{_0})^2\D h_0{^0}=\cO_4\, ,\qquad
  \D b^a{_a}+\dsv^2\D h_a{^a}=\cO_4\, ,\nn
\eea
where we used the notation $\D X\equiv X-\tilde X$. One should also
note that the SdS solution automatically satisfies all the
constraints.

\prg{1.} We begin with the constraint $\cH_\a$, written in the form
\bea
\cH_\a&&=\pi_a{^\b}T^a{}_{\a\b}-b^a{_\a}\pd_\b\pi_a{^\b}+\cO_5\nn\\
&&=\pi_a{^\b}\pd_\a b^a{_\b}-\pd_\b(b^a{_\a}\pi_a{^\b})+\cO_5\, .\nn
\eea
Then, using the formula
$$
\D(\pi_a{^\b}\pd_\a b^a{_\b})=-\frac{4a}{\ell}\D(\pd_\a J)+\cO_5\, ,
$$
one finds
$$
\cH_\a=-\D\left(\pd_\g\cP_\a{^\g}\right)+\cO_5\, .
$$
After noting that $\tilde \cP_\a{^\g}=0$, we finally obtain
\be
\cH_\a=-\pd_\g\cP_\a{^\g}+\cO_5\, .                      \lab{C.1}
\ee

\prg{2.} Going now to $\cH_{ac}$ and using the same technique, we find
\bsubeq
\bea
\cH_{ac}&&=2\left(\dspi_{[a}{^\a}\D b_{c]\a}
  +\D\pi_{[a}{^\a}\dsb_{c]\a}\right)+\pd_\a\pi_{ac}{^\a}+\cO_4\nn\\
 &&=2\left(\dspi\D b_{[ca]} +\dsv\D\pi_{[ac]}\right)
                   +\pd_\a\pi_{ac}{^\a}+\cO_4 \, .       \lab{C.2a}
\eea
In particular,
\be
\pi_{[ac]}+\frac{4a}{\ell}fb_{[ac]}=\cO_3\, .            \lab{C.2b}
\ee
\esubeq
If we further note that
\bea
\cP_{[ac]}&=&\dspi_{e[c}\D b^e{}_{a]}
             +\D\pi_{e[c}\dsb^e{}_{a]}+\cO_4\nn\\
&=&\dspi\D b_{[ca]}+\dsv\D\pi_{[ac]}+\cO_4\, ,\nn
\eea
we verify the final expression \eq{4.2b}.

\prg{3.} The constraint $\cH_{0c}$, after discarding $\cO_3$ terms,
reads
$$
\cH_{0c}=\pi_0{^\a}\bar b_{c\a}-\bar\pi_c{^\a}b_{0\a}
          +\pd_\g\pi_{0c}{^\g}+\cO_3\, ,
$$
which leads to the result \eq{4.2c}.

Using the fact that
$\pi_{0c}{^\g}=4\l_{0c}{}^{0\g}+\hcO=-2a\tilde H_{0c}^{0\g}+\cO_2$,
we find
\be
\pi_{0\b}+\frac{4a}{\ell}fb_{0\b}=
            8af\psi_{,\b}+\cO_3\, .                    \lab{C.3}
\ee
Since the divergence of $\pd_\g\psi\sim n^\g/\r^2$ vanishes, the
preceding relation implies
\be
\pd^\b\left(\pi_{0\b}+\frac{4a}{\ell}fb_{0\b}\right)=\cO_4\, .
                                                       \lab{C.4}
\ee

\prg{4.} Using the general definition, eq. (A1.4b) in Ref. \cite{5},
the dynamical component of the Hamiltonian takes the form
$$
N\cH_\ort=\fr{1}{2}\pi_a{^\a}T^a{}_{0\a}
           +2a\L b-\pd_\b\pi_0{^\b}+\cO_4\, .
$$
The substitution $X=\tilde X+\D X$ yields
$$
N\cH_\ort=\frac{f}{2\ell}\D\pi_a{^a}
 -\frac{2a}{\ell}\dsv^2(\pd_0\D b^c{_c}-\pd_c\D b^c{_0})
 +2a\L \D b-\pd_\b\pi_0{^\b}+\cO_4\, .
$$
In order to eliminate the velocity $\pd_0\D b^c{_c}$ we use the
relation
$$
b^0{_0}\pi_a{^\a}b^a{_\a}=-4aJh_c{^\g}T^c{}_{0\g}+\cO_4\, ,
$$
which follows from the definition of $\pi_a{^\a}$. It leads to
$$
\frac{f}{\ell}\left(\frac{1}{2}\D\pi_a{^a} +\frac{8a}{\ell}\dsv\D
b^c{_c}\right)=2a\L\D b-\frac{2a}{\ell}\dsv^2
    \left(\pd_0\D b^c{_c}-\pd_c\D b^c{_0}\right)+\cO_4\, .
$$
Substituting $\pd_0\D b^c{_c}$ from this equation into $N\cH_\ort$
leads to
\be
N\cH_\ort=\frac{f}{\ell}\left(\D\pi_a{^a}
   +\frac{8a}{\ell}\dsv\D b^c{_c}\right)-\pd_\b\pi_0{^\b}+\cO_4\, .
                                                        \lab{C.5}
\ee
Since $\pd_\b\pi_0{^\b}=\cO_3$, we have
\be
\D\pi_a{^a}+\frac{8a}{\ell}f\D b^c{_c}=\cO_3\, .        \lab{C.6}
\ee
The constraint $N\cH_\ort$ can be rewritten in another useful form:
$$
N\cH_\ort=\frac{1}{\ell}\D\cP_\a{^\a}-\pd_\b\pi_0{^\b}+\cO_4\, .
$$
Now, taking into account that $\tilde\cP_\a{^\a}=0$ and
$N=1 +\cO_1$, we find
\be
\cH_\ort=\frac{1}{\ell}\cP_\a{^\a}
                          -\pd_\b\pi_0{^\b}+\cO_4\, .   \lab{C.7}
\ee
As a consequence, $\cP_\a{^\a}=\cO_3$.

\section{Lagrangian form of the conserved charges}

In this Appendix, we derive equations \eq{6.8} representing the
Lagrangian form of the conserved charges. Most of the needed
formulas are taken from Appendix D of Ref. \cite{22}.

Using the defining relation for the momentum variables $\p_i{^\m}$
($\p_i{^\m}=-4bh^{j\m}\b_{ij}{^0}$), we straightforwardly
express $\p_0{^\g}$ and $\cP_\a{^\g}$ in terms of Lagrangian
variables:
\bsubeq\lab{D.1}
\be
\p_0{^\g} = 2ah^{a\g}\left(\pd_\b H^{\b 0}_{a0}
            + \frac{2}{\ell}bh_a{^0}\right) + \cO_3\,,   \lab{D.1a}
\ee
\bea
\cP_\a{^\g}=
&-& \frac{a}{\sqrt{-g}}g_{\a\b}\left(\dot H^{\b 0}_{c0}H^{c0\g 0}
    +\dot H^{\g 0}_{c0}H^{c0\b 0}\right)
    +\frac{4a}{\ell}J\d^\g_\a                            \nn\\
&+& a\eta_{\a\b}\left(H^{\b\g 0\d}-fH^{\b 0\g\d}
    -fH^{\g 0\b\d}\right)_{,\,\d}                        \nn\\
&+& a(2B-1)g_{\a\b}J\TA{^{\b\g 0}} + \cO_4\,.            \lab{D.1b}
\eea
\esubeq
In the above formulas, $H_{ij}^{\m\n}=H_{ij}{}^{\m\n}$,
$H^{\m\n\l\r}=\d^\m_i\d^\n_j H^{ij\l\r}$, and $\TA{^{\m\n\l}}=
h^{i\m}h^{j\n}h^{k\l}\TA_{ijk}\,$. We want to compare these
with the GR expressions. Rewriting the well known superpotential of
Landau and Lifshitz \eq{6.7b} in terms of tetrad fields, we
obtain
\bsubeq\lab{D.2}
\be
h^{00\g}=2abh^{a\g}\pd_\b H^{0\b}_{0a}
         +\cO^{[\a\g]}{}_{,\,\a} + \cO_3\,,              \lab{D.2a}
\ee
\be
h^{\a 0\g}=
-a\left(\dot H^{\a 0}_{c0}H^{c0\g 0}
        +\dot H^{\g 0}_{c0}H^{c0\a 0}\right)
+af\left(H^{\a\d 0\g}-fH^{\g 0\a\d}\right)_{,\,\d}
+\cO_4\,.                                                \lab{D.2b}
\ee
\esubeq
Now, comparing \eq{D.1} and \eq{D.2}, and using the quantity $\hat
h^{\m 0\l}$ defined in \eq{6.7b} to shorten writing, we arrive at
\bsubeq\lab{D.3}
\be
\p_0{^\g}=\hat h^{00\g}
    +\frac{2a}{\ell}f^2\left(2h^{\g 0}-fg^{0\g}\right)
    +\cO^{[\a\g]}{}_{,\,\a} + \cO_3\, ,                  \lab{D.3a}
\ee
\be
\cP_\a{^\g}=\hat h_\a{}^{0\g}
+a\eta_{\a\b}\left(H^{\b\g 0\d}-H^{\b\d 0\g}
                -fH^{\b 0\g\d}\right)_{,\,\d}+\cO_4\,,   \lab{D.3b}
\ee
\esubeq
where $\hat h_\m{}^{0\l}=g_{\m\n}\hat h^{\n 0\l}$, and
$h^{\m\n}\equiv\d^\m_i h^{i\n}$. The second term in \eq{D.3b} is
$\cO_3$, and we straightforwardly verify the Lagrangian expression
for the conserved momentum. To obtain the same for the Killing
energy, we first notice that
$$
\eta_{\a\b}\left(H^{\b\g 0\d}-H^{\b\d 0\g}
                -fH^{\b 0\g\d}\right)_{,\,\d}
=f^2\left[\left(2h^{\b 0}-fg^{0\b}\right)\d^\g_\a
-\mbox{ $(\b\leftrightarrow\g)$ }\right]_{,\,\b}+\cO_4 \,.
$$
Using this in $\p_0{^\g}-x^\a\cP_\a{^\g}/\ell$, we verify the second
formula of \eq{6.7a}, and consequently, the Lagrangian expression
for the conserved Killing energy.

When applied to the angular momentum, the same procedure gives
$$
x_\a\cP_\b{^\g}-x_\b\cP_\a{^\g}=
x_\a\hat h_\b{}^{0\g}-x_\b\hat h_\a{}^{0\g}+
\frac{f}{(-g)}\psi^\g{}_{\a\b 0}-
2aH^{0\g}_{\a\b}+\cO^{[\d\g]}_{\a\b\,,\,\d}+\cO_3\,,
$$
which leads to the Lagrangian expression for the angular momentum,
as given in \eq{6.8}.

\end{document}